\documentclass[floatfix,twocolumn,aps,prd,showpacs,amsmath,amssymb]{revtex4}

\usepackage{epsfig}
\usepackage{color}

\begin{document}

\title{Scattering of massless scalar waves by Reissner-Nordstr\"om
black holes}
\author{Lu\'\i s C. B. Crispino}
\email{crispino@ufpa.br}
\affiliation{Faculdade de F\'\i sica, Universidade Federal do
Par\'a, 66075-110, Bel\'em, PA, Brazil}

\author{Sam R. Dolan}
\email{sam.dolan@ucd.ie}
\affiliation{School of Mathematical Sciences, University College
Dublin, Belfield, Dublin 4, Ireland}

\author{Ednilton S. Oliveira}
\email{ednilton@fma.if.usp.br}
\affiliation{Instituto de F\'\i sica, Universidade de S\~ao Paulo, 
CP 66318, 05315-970, S\~ao Paulo, SP, Brazil}

\date{\today}
\begin{abstract}

We present a study of scattering of massless planar scalar waves by a 
charged non-rotating black hole. Partial wave methods are applied to 
compute scattering and absorption cross sections, for a range of 
incident wavelengths. We compare our numerical results with semi-classical 
approximations from a geodesic analysis, and find excellent agreement. 
The glory in the backward direction is studied, and its properties are 
shown to be related to the properties of the photon orbit. The effects of 
black hole charge upon scattering and absorption are examined in detail. 
As the charge of the black hole is increased, we find that the absorption 
cross section decreases, and the angular width of the interference fringes 
of the scattering cross section at large angles increases. 
In particular, the glory spot in the backward direction becomes wider. 
We interpret these effects under the light of our geodesic analysis.

\end{abstract}

\maketitle

\section{Introduction}

Almost a century ago, Schwarzschild discovered a pleasingly simple
exact solution to Einstein's gravitational field equations.  Ever
since, exact solutions have been cherished by theoretical physicists
as islands of refuge~\cite{ExactSolutions}. That is to say, natural
harbors, from which the choppy waters of the non-linear dynamical
theory may be safely explored.

Exact black hole solutions are both elegant and simple. 
Members of the Kerr-Newman family depend on just three 
numbers: mass $M$, charge $Q$ and angular momentum $J$~\cite{MTW}. 
Uniqueness and stability proofs suggest that these simple 
stationary spacetimes arise as the generic final outcomes 
from complicated dynamical processes such as stellar collapse~\cite{Hawking-Ellis}. 

To examine dynamics, one may try perturbing a black hole away 
from its stationary configuration. The interaction of fields 
with Kerr-Newman black holes is of relevance to questions about 
formation, stability, and gravitational wave emission. 
For example, is now well established that 
\emph{black holes have no hair}; in other words, all 
long-ranged classical fields (`hair') must decay away~\cite{Frolov-Novikov}.  

In the 70s and 80s, significant effort was devoted to the study 
of the scattering and absorption of planar waves that impinge upon 
black holes in vacuum (cf., e.g., Ref.~\cite{FHM} and references therein). 
This subject has also received attention in recent years 
(e.g., see Refs.~\cite{Glampedakis-Andersson-2001, DolanF, 
Dolan-2006, CHMO, Dolan-2008}). In the standard scenario, 
authors consider a black hole irradiated by a long-lasting 
monochromatic plane wave of frequency $\omega$ which is 
incident from infinity. Flux is absorbed and scattered, and, 
if the wave has intrinsic spin, polarized. 
The resulting scattering pattern may be interpreted as the 
signature of the black hole. Its features depend primarily upon the 
dimensionless coupling $\omega M$. The large-angle scattering pattern, 
in particular the so-called \emph{glory} in the backward direction, 
is inextricably linked to the near-horizon geometry of the hole. 
It is conceivable that such patterns may one day be observed 
experimentally at gravitational wave detectors.

Although Reissner-Nordstr\"om black holes have not received 
the same degree of attention as Schwarzschild and Kerr black holes, 
some effort has been devoted to study the emission and absorption 
properties of charged black holes. 
In 1977, Page~\cite{PageIII} considered 
the Hawking emission rates from a nonrotating black hole of small 
charge, calculated for electrons and muons and their antiparticles.
Absorption properties of massive scalars by Reissner-Nordstr\"om black 
holes were analyzed by Jung, Kim, and Park~\cite{JKP}. 
The absorption and emission spectra of higher-dimensional static charged 
black holes have been computed by Jung and Park both in the brane 
and in the bulk for the massless scalar field~\cite{JP}.
The electromagnetic absorption cross section of Reissner-Nordstr\"om black holes
has been studied by two of the present authors~\cite{CO}.
Notwithstanding, to the best of our knowledge, there are no previous 
works devoted to planar wave scattering by 
Reissner-Nordstr\"om black holes in the literature.
The present paper is dedicated to the study of scattering 
and absorption of massless scalar waves by static charged
black holes in four dimensions.

The Reissner-Nordstr\"om spacetime line element is given by
\begin{equation}
 ds^{2}= f(r)dt^{2} - [f(r)]^{-1}dr^2 - r^{2}\left(d\theta^2+\sin^2
\theta d\phi^2 \right),
 \label{rn}
\end{equation}
where $f(r)=(1-r_{+}/r)(1-r_{-}/r)$ with $r_{\pm}=M \pm
\sqrt{M^2 - Q^2}$. 
We use natural units with $c = G = 1$ and the metric
signature $(+ - - -)$.

In this work, we exhibit results for three different absolute values of the
black hole charge, namely: $Q=0$, $|Q|=0.5M$, and $|Q|=M$. Here, $Q=0$ is the
Schwarzschild case which was investigated in Refs.~\cite{Sanchez,
  Andersson}, $|Q|=0.5M$ is a typical Reissner-Nordstr\"om black
hole example, and $|Q|=M$ is the extreme Reissner-Nordstr\"om black
hole case. Our
formalism can be used to obtain results for arbitrary values of the
ratio $q\equiv |Q|/M$, in the interval $0 \le q \le 1$.

The remainder of the paper is organized as follows: In Sec.~\ref{sec-CA}
we consider the geodesics of the  
Reissner-Nordstr\"om spacetime. The partial wave approach is 
outlined in Sec.~\ref{sec-partialwaveanalysis}, 
where we give expressions for the
massless scalar field, and the absorption and scattering cross sections.
Our numerical results are presented in Sec.~\ref{sec-results}.
We conclude with some final remarks in Sec.~\ref{sec-finalremarks}.

\section{Classical Analysis\label{sec-CA}}

Here we analyze geodesics in the Reissner-Nordstr\"om spacetime. 
The key results obtained in this section are used to check the 
validity of our numerical results, obtained from the partial
wave scattering analysis of Sec.~\ref{sec-partialwaveanalysis}.

The geodesics of the Reissner-Nordstr\"om spacetime can be found by using
Eq.~(\ref{rn}) to write
\begin{equation}
 \dot{s}^{2} = f(r) \dot{t}^{2} - [f(r)]^{-1} \dot{r}^{2} - r^{2}
 \left(\dot{\theta}^{2} +  \dot{\phi}^{2} \sin^{2}\theta \right) =
 \kappa \, ,
 \label{rnk}
\end{equation}
where the ``.'' denotes the derivative with respect to an affine parameter.
For massive particles we have $\kappa=1$, and for massless particles we have $\kappa=0$. 

The orbit equation for massless particles is \cite{Chandrasekhar}
\begin{equation}
 \left(\frac{du}{d\phi} \right)^{2} = \frac{1}{b^{2}} - u^{2} +
2Mu^{3} - Q^{2} u^{4},
 \label{Geo1}
\end{equation}
where $u=1/r$ and $b$ is the impact parameter.
By integrating Eq.~(\ref{Geo1}) we obtain the deflection angle
\begin{equation}
\Theta(b) = \frac{4}{\sqrt{Q^2(u_3 - u_1)(u_2 - u_0)}} 
\left[ K(k) - F(z, k) \right] - \pi \,,
 \label{scatt_angle}
\end{equation}
where $F(z, k)$ and $K(k)$ are the incomplete and complete elliptic 
integrals of first kind~\cite{Abramowitz-Stegun}, respectively, with
$$ k^2= \frac{(u_3 - u_2)(u_1 - u_0)}{(u_3 - u_1)(u_2 - u_0)},
$$
and
$$
z = \left[ -\frac{u_0 (u_3 - u_1)}{u_3 (u_1 - u_0)}  \right]^{1/2}.
$$
Here, $u_{0}$, $u_{1}=1/r_{\text{min}}$, $u_{2}$ and $u_{3}$ are roots of the
right hand side of Eq.~(\ref{Geo1}), and $r_{\text{min}}$ is the radius of closest approach. 
For scattering geodesics, the roots obey the inequalities
$u_0 < 0$ and $u_3 > u_2 \ge u_1 > 0$. 
(An analysis of the scattering of null geodesics on the 
Reissner-Nordstr\"om spacetime is presented in Appendix \ref{Glory_approx}. 
A more extensive study of geodesics on black hole spacetimes may be 
found in \cite{Slezakova}, for example.)

By differentiating Eq.~(\ref{Geo1}) we get
\begin{equation}
 \frac{d^{2}u}{d\phi^{2}} + u = 3Mu^{2} - 2Q^{2} u^{3}.
 \label{Geo2}
\end{equation}

\begin{figure}
 \centering
 \includegraphics[scale=1]{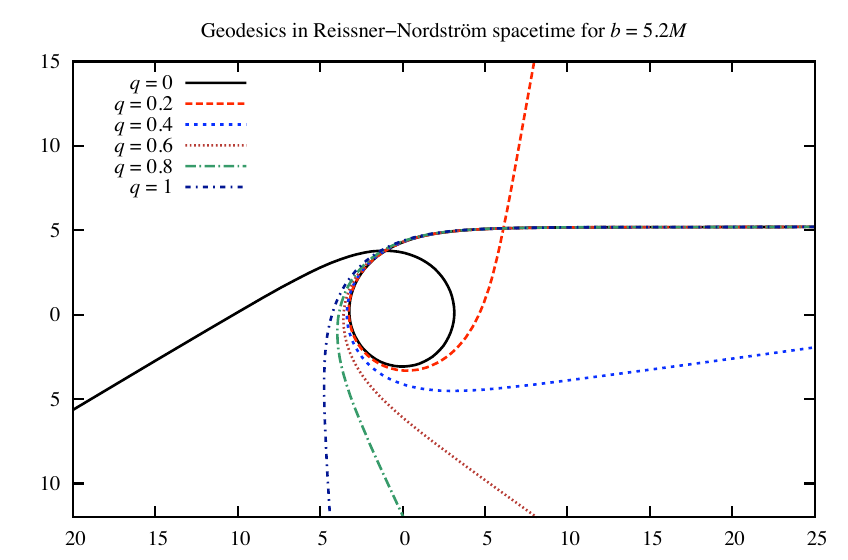}
 \caption{Geodesics in the Reissner-Nordstr\"om spacetime for different
values of the ratio $q=|Q|/M$. Here, the impact parameter has been 
chosen to be $b=5.2M$. We see
that the black hole has a stronger influence in the particle
trajectory for smaller values of $q$.}
 \label{Geo}
\end{figure}

We solved Eqs.~(\ref{Geo1}) and~(\ref{Geo2}) numerically to examine
how the black hole charge influences geodesics in Reissner-Nordstr\"om 
spacetime, and compared with the Schwarzschild case. 
In Fig. \ref{Geo} we compare the 
geodesics on different Reissner-Nordstr\"om spacetimes. 
The mass of the hole $M$ is fixed but the charge-to-mass 
ratio $q = |Q|/M$ is varied. We find that, for a fixed impact parameter $b$, 
a larger ratio $q$ leads to a smaller deflection angle $\Theta$.

Using Eqs.~(\ref{Geo1}) and~(\ref{Geo2}) we may derive 
an analytical approximation to the scattering cross section for small angles. 
Considering the weak field limit, 
the deflection angle is found to be~\cite{Eiroa, Bhadra, Sereno}:
\begin{equation}
 \Theta (b) \approx \frac{4M}{b} + \frac{3\pi}{4} 
 \left( 5 - q^2 \right)\frac{M^2}{b^2} \,. \\
 \label{csa}
\end{equation}
Note that for large impact parameters we obtain 
$\Theta (b) \approx 4M/b$, 
which is Einstein's deflection angle~\cite{Wald}.

The classical differential scattering cross section is given by
\begin{equation}
 \left. \frac{d\sigma_{sc}}{d\Omega} \right|_{cl} = \frac{b}{\sin\theta} \left|\frac{db}{d
 \theta} \right|.
 \label{cscs}
\end{equation}
From Eqs.~(\ref{csa}) and~(\ref{cscs}) we conclude that the
classical differential scattering cross section for small angles is
\begin{equation}
 \left. \frac{d\sigma_{sc}}{d\Omega} \right|_{cl} \approx 
 \frac{16M^2}{\theta^4}+\frac{15\pi M^2}{4\theta^3}
 - \frac{3\pi Q^2}{4\theta^3}.
 \label{wfscs}
\end{equation}
We see that, in the weak field limit, the presence of the black hole 
charge does not modify the dominant term neither in the 
deflection angle nor in the scattering cross section. 
We can thus conclude that in the high-frequency (short-wavelength) 
limit the differential scattering cross section for small angles 
must be approximately independent of the black hole charge. 

The presence of an unstable photon orbit at $r = r_c$ (see Appendix \ref{Glory_approx}) 
means that, in theory, geodesics may be deflected through any angle. 
This property, together with the axial symmetry of our plane-wave scattering scenario, 
implies that a \emph{glory} will be present. 
Just as in optics, a glory is a bright spot or halo arising in the scattered 
intensity in the antipodal direction. The intensity and size of the spot or 
halo depends on the wavelength of the incident perturbation, leading to chromatic effects. 
The magnitude and size of the bright spot may be estimated using the approximation 
derived by Matzner \emph{et al.} \cite{MMNZ}
\begin{equation}
 \left.\frac{d\sigma_{sc}}{d\Omega}\right|_{\theta\approx\pi} \approx
 2\pi\omega b_{g}^{2} \left|\frac{db}{d\theta}\right|_{\theta=\pi}
 [J_{2s}(\omega b_{g} \sin\theta)]^2\,.
 \label{Gen glory}
\end{equation}
Here, $b_{g}$ is the impact parameter that corresponds to a deflection angle of 
$\pi$, $J_{2s}(x)$ is a Bessel function of the first kind (of order $2s$), 
and $s$ is the spin of the field ($s = 0$ for the scalar wave). 
We recall that Eq.~(\ref{Gen glory}) is 
an approximation valid at high frequencies ($\omega M \gg 1$), 
for angles close to the backward direction ($\theta \approx \pi$).

The value of $b_{g}$ can be obtained $(i)$ by numerically solving the 
orbit equation (\ref{Geo1}), or $(ii)$ by considering an analytical 
approximation of the deflection angle, valid for impact parameters 
close to the critical one ($b \approx b_c$). 
We compare the results of methods $(i)$ and $(ii)$ in Appendix
\ref{Glory_approx}. We find that there is a significant
difference between these two approaches. The approximate approach $(ii)$ 
suggests that the glory peak intensity will decrease monotonically as the 
black hole charge-to-mass ratio is increased. This is not supported by the 
numerical approach $(i)$, however. We find that the peak intensity decreases, 
reaches a minimum, and increases again, as $q$ increases. 
This demonstrates that the approximate method 
(based on the `Darwin approximation' \cite{Darwin}) is not sufficiently 
accurate for our purposes. 

We combined method $(i)$ with the glory approximation (\ref{Gen glory}), 
to estimate the magnitude and width of the glory. We found
\begin{equation}
 \left.\frac{d\sigma_{sc}}{d\Omega}\right|_{\theta\approx\pi}
 \approx \mathcal{A}(q) [J_{0}(b_g(q) \omega \sin\theta)]^2,
\label{Glory 0}
\end{equation}
where $\mathcal{A}(q)/(\omega M^3)  = \{30.75, 29.73, 28.87\}$ and $b_g(q) / M = 
\{5.36, 5.14, 4.30 \}$ for $q = 0, 0.5, 1$, respectively. 
In Sec.~\ref{sec-results} we check the scattering cross 
section obtained via the partial wave method against this semi-classical prediction.

\section{Partial Wave Analysis\label{sec-partialwaveanalysis}}

\subsection{Massless Scalar Field Equation}

In curved spacetimes the equation for the minimally-coupled 
massless scalar field is
\begin{equation}
 \nabla^{\mu} \nabla_{\mu}\Phi=0\,.
 \label{KG eq}
\end{equation}
For $r>r_{+}$, the Reissner-Nordstr\"om spacetime, 
which is spherically symmetric, has a global
timelike Killing field, $\partial_{t}$. Hence, we may write
\begin{equation}
 \Phi = \frac{\psi_{\omega l}(r)}{r} Y_{lm} (\theta, \phi) e^{-i
 \omega t}.
 \label{Separation}
\end{equation}
Here, $Y_{lm}(\theta, \phi)$ are the scalar spherical harmonics. 
The radial solutions $\psi_{\omega l}$ can be expressed in terms of two 
independent sets of modes: one incoming from the past white-hole horizon 
${\cal H}^{-}$ and other incoming from the past null
infinity ${\cal J}^{-}$ (see, e.g., Ref.~\cite{CSM} for more detail). 
Here we are dealing with scattering of waves by black holes, 
so that we need only to consider those modes incoming from 
${\cal J}^{-}$. 

The equation for $\psi_{\omega l}$ can be written as
\begin{equation}
 f(r) \frac{d}{dr} \left[f(r) \frac{d}{dr} \psi_{\omega l}(r) \right]
 + \left[\omega^2 - V_{eff}(r) \right] \psi_{\omega l} (r) = 0,
 \label{Radial eq}
\end{equation}
where the effective potential is given by
\begin{equation}
 V_{eff}(r) = f(r) \left[\frac{1}{r}\frac{df(r)}{dr} +
 \frac{l(l+1)} {r^2} \right].
 \label{Veff}
\end{equation}

To better treat the solution of Eq.~(\ref{Radial eq}) in the asymptotic
limits, we introduce the tortoise coordinate $x$ defined by
\begin{equation}
 \frac{d}{dx} = f(r) \frac{d}{dr},
 \label{Tortoise diff}
\end{equation}
or, in integral form,
\begin{equation}
 x = r + \frac{r_{+}^{2}}{r_{+}-r_{-}}\ln{\left|\frac{r}{r_{+}}-1
 \right|} - \frac{r_{-}^{2}}{r_{+} - r_{-}}
 \ln{\left|\frac{r}{r_{-}}-1 \right|} + C,
 \label{Tortoise coord}
\end{equation}
where $C$ is an integration constant. 
Our numerical results for the scattering cross section are independent 
of the choice of $C$, and we have set $C=0$.

In terms of the tortoise coordinate, the radial equation (\ref{Radial eq})
may be written as 
\begin{equation}
 \frac{d^2}{dx^2} \psi_{\omega l}(x) + \left[\omega^2 - V_{eff}(x)
 \right] \psi_{\omega l} (x) = 0.
 \label{Radial eq x}
\end{equation}

\begin{figure}
 \centering
 \includegraphics[scale=1]{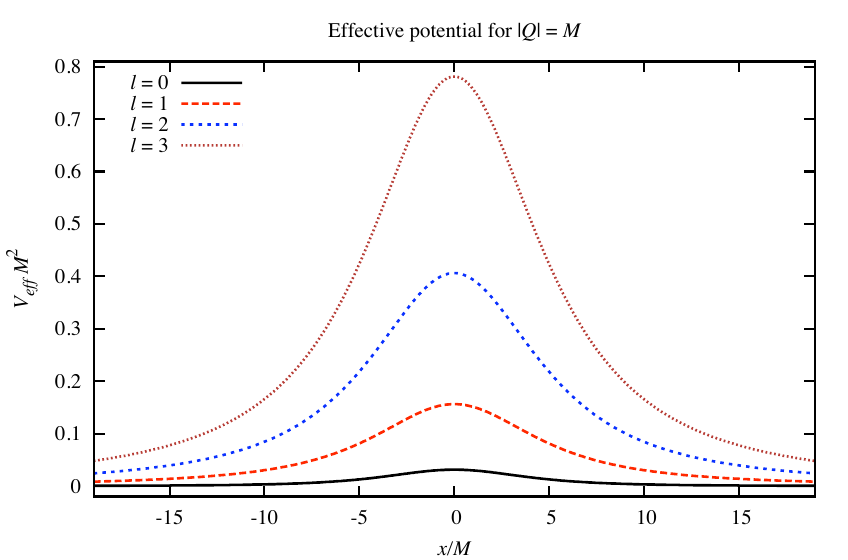}
 \caption{The effective potential for the extreme Reissner-Nordstr\"om black
hole as a function of the tortoise coordinate~(\ref{Tortoise coord}), 
plotted for different choices of $l$.}
 \label{px-qm}
\end{figure}

In Fig.~\ref{px-qm} we plot the effective potential as a function
of the tortoise coordinate for an extreme Reissner-Nordstr\"om black
hole.

\begin{figure}
 \centering
 \includegraphics[scale=1]{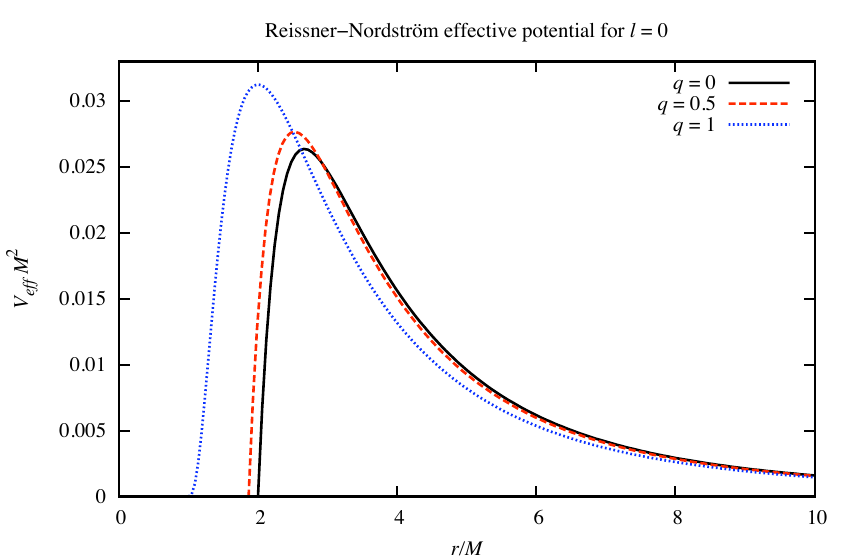}
 \caption{The effective potential given by Eq.~(\ref{Veff}) with $l=0$ 
 is plotted for $q=0$ (solid line), $q=0.5$ (dashed line), and $q=1$ (dotted line). 
 The effective potential goes to zero at
 the event horizon and at infinity. As we can see, the maximum of
the potential increases as the black hole charge increases.}
 \label{pr}
\end{figure}

In Fig.~\ref{pr} we compare the effective potential, with $l=0$, for
the three cases: $Q=0$, $|Q|=0.5M$, and $|Q|=M$. As we can see, the
effective potential goes to zero as $r\rightarrow r_{+}$ and as
$r\rightarrow \infty$, for all cases. The height of the 
effective potential barrier increases with the charge-to-mass ratio $q$.

For $r \gg r_{+}$, we have
\begin{equation}
 \psi_{\omega l}(x) \approx \omega x\left[ (-i)^{l+1}A_{\omega
 l}^{in}
 h_{l}^{(1)*}(\omega x) + i^{l+1} A_{\omega l}^{out} h_{l}^{(1)} 
 (\omega x)
 \right] ,
 \label{psi for r>>r+}
\end{equation}
where $h_{l}^{(1)} (x)$ are the spherical Bessel functions of the
third 
kind~\cite{Abramowitz-Stegun}, and $A_{\omega l}^{in}$ 
and $A_{\omega l}^{out}$ are complex constants.

Now, recalling that $h_{l}^{(1)}(x) \approx (-i)^{l+1}e^{ix}/x$ as
$x \rightarrow \infty$, and using that the effective potential goes to
zero as $x \rightarrow -\infty$, we get
\begin{equation}
 \psi_{\omega l} (x) \approx \left\{
 \begin{array}{lr}
  A_{\omega l}^{tr} e^{-i\omega x} & (x \rightarrow -\infty), \\
  A_{\omega l}^{in} e^{-i\omega x} + A_{\omega l}^{out} e^{i\omega x}
  & (x \rightarrow
  \infty),
 \end{array} \right.
 \label{psi at infinity}
\end{equation}
with the relation $|A_{\omega l}^{in}|^{2} = |A_{\omega l}^{out}|^2 +
|A_{\omega l}^{tr}|^2$ satisfied.

\subsection{Absorption Cross Section}

The total absorption cross section can be written as
\begin{equation}
 \sigma_{abs} = \sum\limits_{l=0}^{\infty} \sigma_{abs}^{(l)},
 \label{tacs}
\end{equation}
where $\sigma_{abs}^{(l)}$ is the partial absorption cross section, namely
\begin{equation}
 \sigma_{abs}^{(l)} = \frac{\pi}{\omega^2}(2l+1) \left(1 - \left|
 \frac{A_{\omega l}^{out}}{A_{\omega l}^{in}} \right|^2 \right).
\end{equation}

From the classical analysis, developed in Sec.~\ref{sec-CA}, the 
geometrical optics (high-frequency) limit of the total absorption 
cross section can be found to be
\begin{equation}
\sigma_{abs}^{hf} = \pi \frac{\left(3M+\sqrt{9M^{2}-8Q^{2}} \right)^{4}}
{8\left(3M^{2} - 2Q^{2} + M\sqrt{9M^{2}-8Q^{2}} \right)}.
\label{GO limit}
\end{equation}
It is easy to check that for $Q=0$ we get from Eq.~(\ref{GO limit}) 
$\sigma_{abs}^{hf} = 27 \pi M^2 = (27/4) \pi r_{+}^{2}$~\cite{Mashhoon1973}.
(We recall that $r_{+} = 2 M$ and $r_{-} = 0 $ in
the Schwarzschild case.)  
For $|Q|=M$ we get $\sigma_{abs}^{hf} = 16
\pi M^2= 16 \pi r_{+}^{2}$. 
(We recall that $r_{+} = r_{-}=M$ in the extreme Reissner-Nordstr\"om case.)

\subsection{Scattering Cross Section \label{pwmscs} }

The phase shifts of the scattered waves are defined by
\begin{equation}
 \exp{[2i\delta_{l}(\omega)]}= (-1)^{l+1} \frac{A_{\omega l}^{out}}{
 A_{\omega l}^{in}}\,.
 \label{ps}
\end{equation}
The scattering amplitude is given by
\begin{equation}
 g(\theta) = \frac{1}{2i\omega} \sum\limits_{l=0}^{\infty} (2l+1)
 \left[e^{2i\delta_{l}(\omega)} - 1 \right] \text{P}_{l}(\cos\theta),
 \label{g}
\end{equation}
and the differential scattering cross section is 
\begin{equation}
 \frac{d\sigma_{sc}}{d\Omega} = \left|g(\theta) \right|^2.
\end{equation}

A selection of our key results for the absorption and scattering
cross sections is presented in the next section.
The numerical method we have used is described in Ref.~\cite{DOC}.
In addition, we have used the method developed in Refs.~\cite{YRW,DolanF}
to improve the numerical convergence of the partial wave series (\ref{g}).

\section{Results\label{sec-results}}

\begin{figure}
 \centering
 \includegraphics[scale=1]{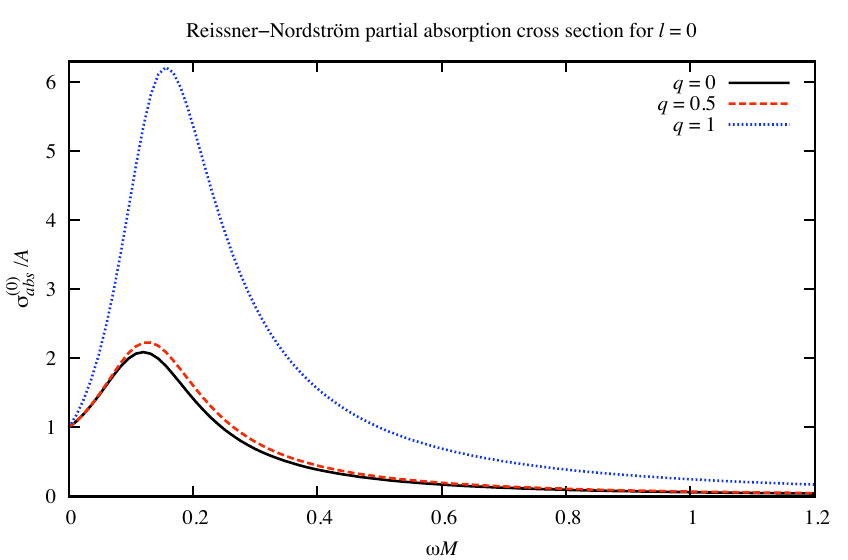}
 \caption{The partial absorption cross section with $l=0$ plotted for 
 Reissner-Nordstr\"om black holes, with 
 $q=0$ (solid line), $q=0.5$ (dashed line), and $q=1$ (dotted line).
 We see that $\sigma_{abs}^{(0)}\rightarrow A$ as $\omega M \rightarrow 0$.}
 \label{pacs}
\end{figure}

In Fig.~\ref{pacs} we plot the partial absorption cross section of
Reissner-Nordstr\"om black holes divided by the
black hole area, $A=4\pi r_{+}^{2}$, for $l=0$, 
and for $q=0$ (Schwarzschild case),
$q=0.5$ (typical Reissner-Nordstr\"om case) and $q=1$ (extreme case).
We see that in the low-frequency limit we have
$\sigma_{abs}^{(0)\,lf} = A$~\cite{Das,Atsushi}. In this limit, the only
nonvanishing contribution to the total absorption cross section
comes from the isotropic mode with $l=0$.

\begin{figure}[!htb]
 \centering
 \includegraphics[scale=1]{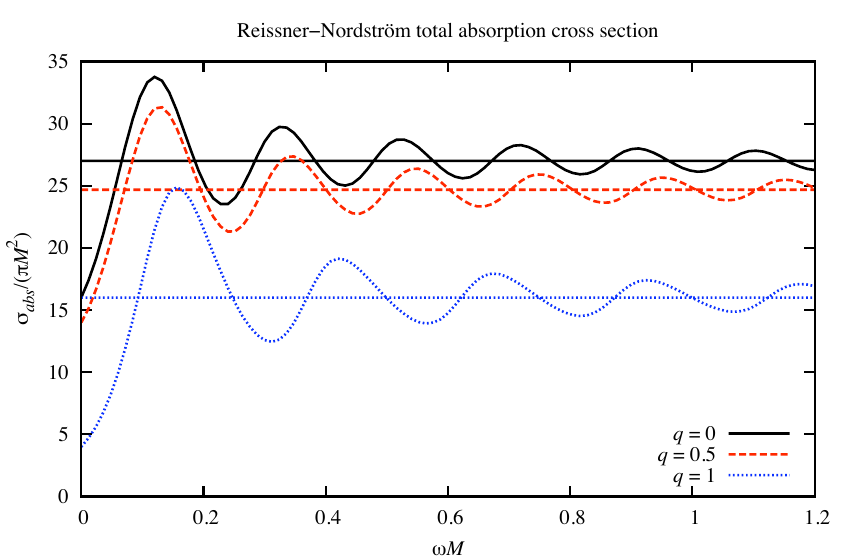}
 \caption{Reissner-Nordstr\"om absorption cross section for $q=0$,
 $q=0.5$, and $q=1$. We see that the
total absorption cross section decreases as the black hole charge increases.}
 \label{tsacs}
\end{figure}

The total absorption cross section of Reissner-Nordstr\"om black holes
is plotted in Fig.~\ref{tsacs}, for the same three choices of the
charge ($q=0, 0.5, 1$).  As we can see, the absorption cross section
decreases as the charge-to-mass ratio increases.  (The same behavior
is observed for the electromagnetic field absorption cross
section~\cite{CO}.)  This is in concordance with the observation that the
height of the effective potential barrier (see Fig. \ref{pr})
increases with the charge-to-mass ratio.  The straight lines in
Fig. \ref{tsacs} show the geometric-optics limit for each case.

\begin{figure}[!htb]
 \includegraphics[scale=1]{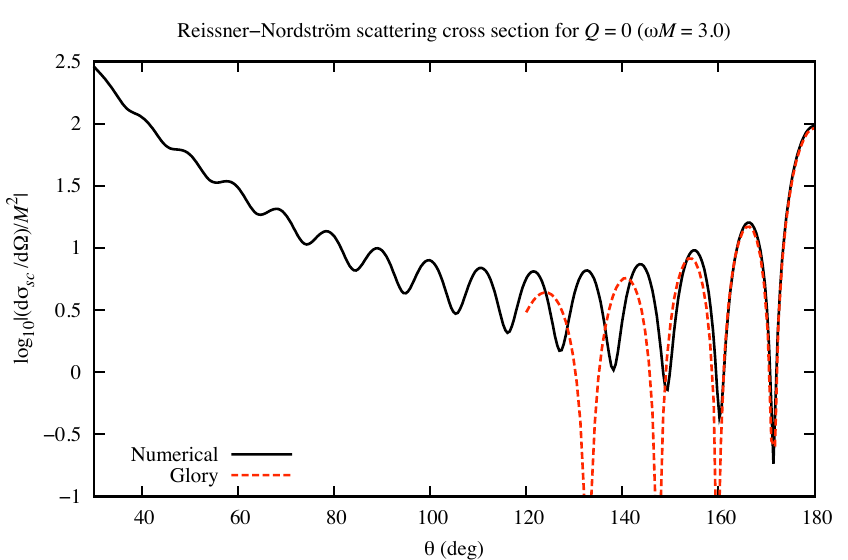}
 \includegraphics[scale=1]{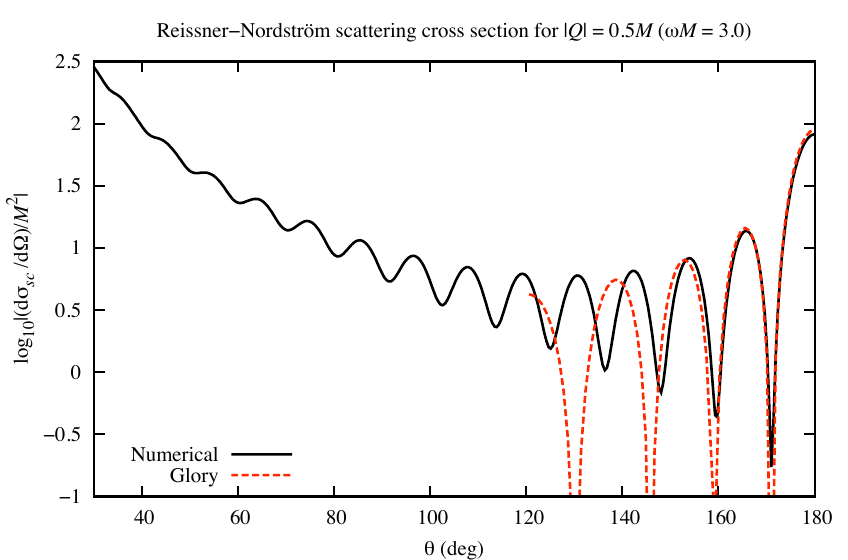}
 \includegraphics[scale=1]{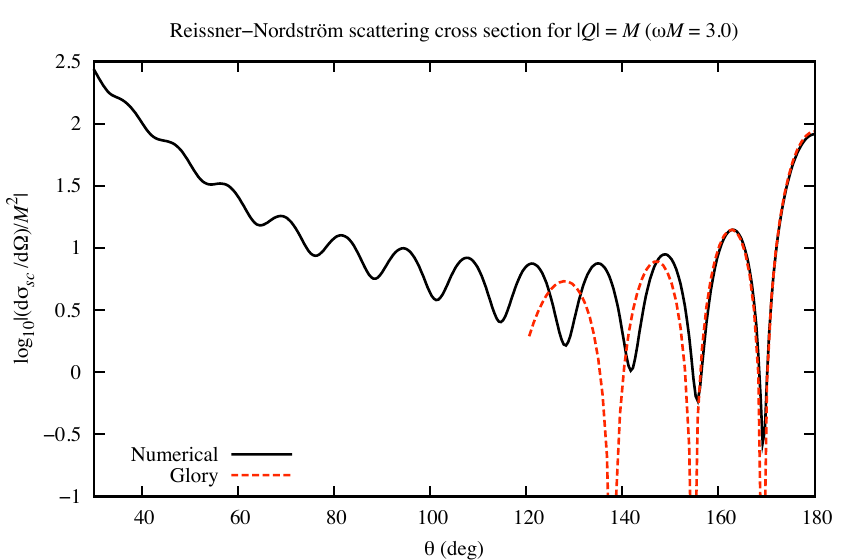}
 \caption{Comparison between the scattering cross section of
Reissner-Nordstr\"om black holes and the glory 
approximation at $\omega M = 3$, for $q=0, 0.5, 1$. 
As we can see, for the three cases, the numerical results 
are in excellent agreement with the glory approximation for 
angles close to $180^{\circ}$.}
 \label{Glory comp}
\end{figure}

In Fig.~\ref{Glory comp} we plot the differential scattering cross
sections of Reissner-Nordstr\"om black holes for the massless
scalar field at $\omega M = 3.0$. 
The values chosen for the black hole charge are again such that
$q=0$, $q=0.5$ and $q=1$. In this figure we also plot the glory scattering
cross sections given in Eq.~(\ref{Glory 0}). 
We find an excellent agreement between the numerical results 
and the glory approximation for $\theta \approx \pi$.

\begin{figure}[!htb]
\includegraphics[scale=1]{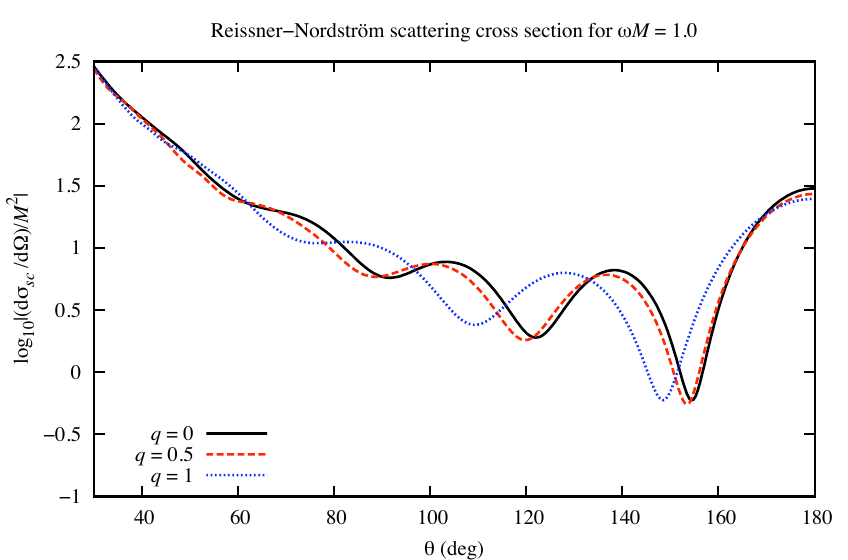}
\caption{Scattering cross section for Reissner-Nordstrom black holes
with charges $q=0$ (solid line), $q=0.5$ (dashed line), and $q=1$ (dotted line), 
for $\omega M=1.0$. 
The width of the glory peak gets wider for bigger values of the 
black hole charge.}
\label{scs-wm1} 
\end{figure}

\begin{figure}[!htb]
\includegraphics[scale=1]{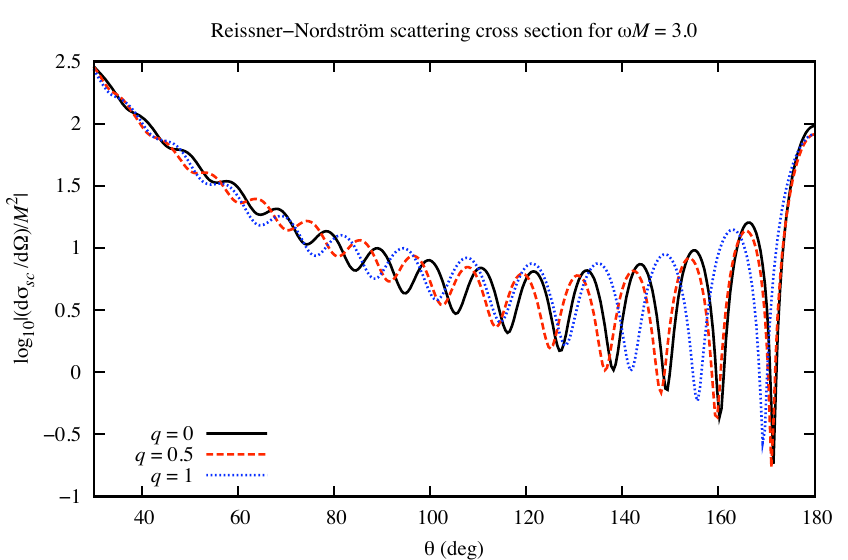}
\caption{Scattering cross section for Reissner-Nordstrom black holes
with charges $q=0$, $q=0.5$ and $q=1$, for $\omega M=3.0$. Here,
as in Fig.~\ref{scs-wm1}, the width of the glory peak increases as
the black hole charge increases.}
\label{scs-wm3} 
\end{figure}

We compare the scattering cross sections in Fig.~\ref{scs-wm1}, for
the same choices of the black hole charge ($q=0, 0.5, 1$),
at $\omega M =1.0$. 
In Fig.~\ref{scs-wm3} we make the same comparison for $\omega M =3.0$. 
We see that, at fixed frequency, the glory peak 
is wider for larger values of the charge-to-mass ratio $q$.
This can be understood by the fact that, from Eq.~(\ref{Gen glory}), 
the glory peak width is 
proportional to $1/(b_{g}\omega)$, and from 
Eq.~(\ref{Glory 0}) we see that $b_{g}$ is smaller for larger values of $q$.

\section{Final Remarks\label{sec-finalremarks}}

In the preceding sections, we have computed absorption and scattering 
cross sections for planar monochromatic massless scalar waves impinging 
upon Reissner-Nordstr\"om black holes. We found that the interaction 
depends on frequency $\omega M$ and charge-to-mass ratio $q = |Q|/M$. 
We developed the formalism needed to obtain scattering and absorption 
cross sections for arbitrary values of $0 \le q \le 1$. 
We showed typical results for three different values of the charge-to-mass 
ratio of the black hole, namely: $q=0$, $q=0.5$, and $q=1$. 

What, then, are the effects of black hole charge upon the scattering and 
absorption of massless scalar waves? Let us summarize. The effect on absorption 
is clear: the absorption cross section decreases as the 
charge-to-mass ratio increases (Fig.~\ref{tsacs}). This is compatible with the fact 
that the horizon area shrinks from $16\pi M^2$ at $q=0$ to $4 \pi M^2$ at $q=1$. 
In our numerical results we have observed that, in the low-frequency limit, 
the absorption cross section tends to the black hole area 
(Fig.~\ref{pacs})~\cite{Das}. This is a general result for the absorption 
cross section of the minimally-coupled massless scalar field in stationary 
black hole spacetimes~\cite{Atsushi}. We have also observed that, 
in the high-frequency limit, the absorption cross sections oscillates 
about the geometric-optics value (Fig.~\ref{tsacs}). 
Similar behavior has previously been observed for the electromagnetic field~\cite{CO}.

The effects of black hole charge upon wave scattering are more subtle. 
By using the weak-field approximation (\ref{csa}), we showed that 
the scattering cross section at small angles $(\theta \approx 0)$ is still 
dominated by the ``Schwarzschild term'' ${16M^2}/{\theta^4}$ 
[cf. Eq.~(\ref{wfscs})]. Black hole charge leads only to a subdominant 
correction term proportional to $Q^2 / \theta^{3}$ at small angles. 
However, the black hole charge does have a significant effect upon the 
cross section observed at large angles (Figs.~\ref{scs-wm1} and~\ref{scs-wm3}). 
We found that the angular width of the so-called spiral scattering oscillations 
increases with $q$. In particular, the glory peak becomes wider as $q$ 
increases (Fig.~\ref{Glory comp}). These effects are related to the 
fact that the radius of the photon orbit shrinks as $q$ increases. 

In principle, highly accurate measurements of, for example, the gravitational 
wave flux scattered by a black hole could one day be used to estimate black 
hole's charge. A more immediate possibility is that scattering and absorption patterns 
may be observed with black hole analogue systems created in the laboratory~\cite{DOC}. 
Even if experimental verification is not forthcoming, we hope that studies 
of wave scattering by black holes will continue to improve our understanding 
of how black holes interact with their environments.

\begin{acknowledgments}
The authors would like to thank Conselho Nacional de Desenvolvimento 
Cient\'\i fico e Tecnol\'ogico (CNPq) for partial financial support. 
S. D. acknowledges financial support from the Irish Research Council 
for Science, Engineering and Technology (IRCSET).
S. D. and E. O. thank the Universidade Federal do Par\'a (UFPA) 
in Bel\'em for kind hospitality. 
L. C. and E. O. would like to acknowledge also partial financial 
support from Coordena\c{c}\~ao de Aperfei\c{c}oamento de Pessoal
de N\'\i vel Superior (CAPES).
\end{acknowledgments}

\begin{appendix}
\section{Analytical approximation for the glory coefficients}
\label{Glory_approx}

Here we derive an approximation to the deflection angle given in
Eq.~(\ref{scatt_angle}), in order to obtain an analytic expression for
the glory impact parameter $b_g$ and its derivative. Our aim is to
estimate the magnitude and width of the glory peak for
Reissner-Nordstr\"om black hole scattering.  Let us begin by finding
the roots of the right hand side of Eq.~(\ref{Geo1}) when the impact
parameter is critical ($b=b_c$). We have
\begin{eqnarray}
\bar{u}_0 &=&  \left[ \left(M + y \right) - 2 \sqrt{M(M+y)} \right] / {4Q^2}\,, \\
\bar{u}_1 &=& u_c = \left( 3M - y \right) / 4Q^2\,,  \\
\bar{u}_2 &=& u_c\,,   \\
\bar{u}_3 &=&  \left[ \left(M + y \right) + 2 \sqrt{M(M+y)} \right] / {4Q^2}\,,
\end{eqnarray}
where $y=\sqrt{9M^2 - 8Q^2}$.
Note that a root is repeated in the critical case: $u_1= u_2 = u_c$. 
The radius of the circular photon orbit is $r_c = 1/u_c$. The critical
impact parameter is
\begin{equation}
b_c = \frac{(3M + y)^2}{\sqrt{8(3M^2 - 2Q^2 + My)} }.
\end{equation}
In the Schwarzschild limit ($Q = 0$) we recover 
$u_0 = -1/6M$, $u_1 = u_2 = 1/3M$, $u_3 = +\infty$ 
and $b_c = \sqrt{27}M$~\cite{Wald}.

When the impact parameter is close to critical ($b \approx b_c$), 
the perturbed roots are 
\begin{eqnarray}
u_0 = \bar{u}_0 && + \mathcal{O}(\delta^2)\,, \\
u_1 = u_c   & - \; u_c \delta \Delta  & + \mathcal{O}(\delta^2)\,,  \\
u_2 = u_c   & + \; u_c \delta \Delta  & + \mathcal{O}(\delta^2)\,, \\
u_3 = \bar{u}_3 && + \mathcal{O}(\delta^2)\,, 
\end{eqnarray}
where $\delta^2 = (b - b_c) / b_c$ and $\Delta^2 = 2 / (6 - u_c^2 b_c^2)$. 

\begin{figure}
\includegraphics[scale=1]{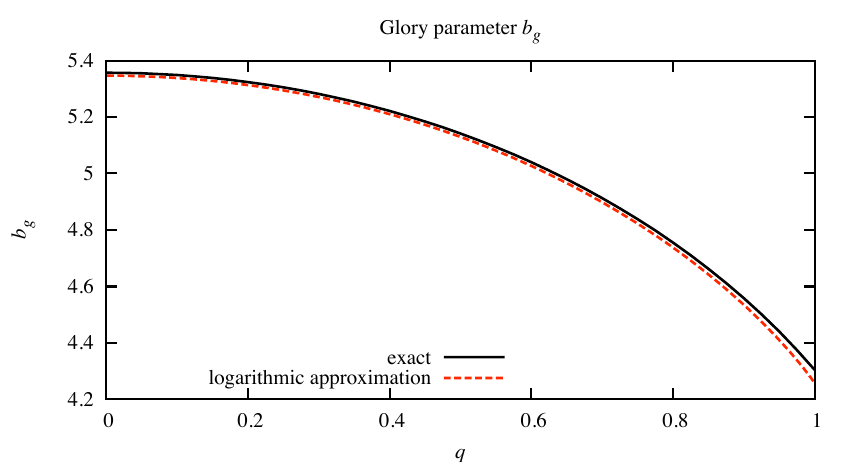}
\includegraphics[scale=1]{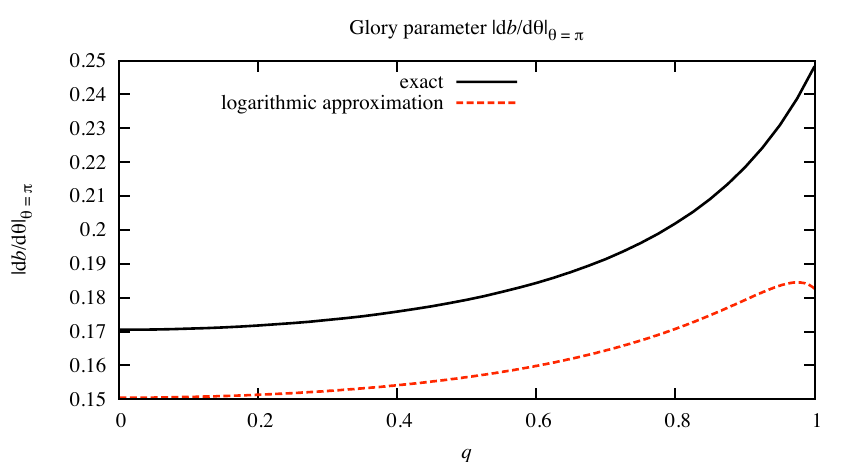}
\caption{Glory parameters $b_g$ and $\left|
{db}/{d\theta}\right|_{\theta = \pi} $
shown as a function of $q$. The plots compare
the approximation Eq.~(\ref{scatt_angle_approx}) (dotted line)
with accurate results from numerical integration (solid line). The
approximation for $\left| {db}/{d\theta}\right|_{\theta = \pi}$
is clearly less
good than the approximation of $b_g$, and the accuracy diminishes
further as $q\rightarrow 1$.}
\label{bgdbdt}
\end{figure}

\begin{figure}
 \includegraphics[scale=1]{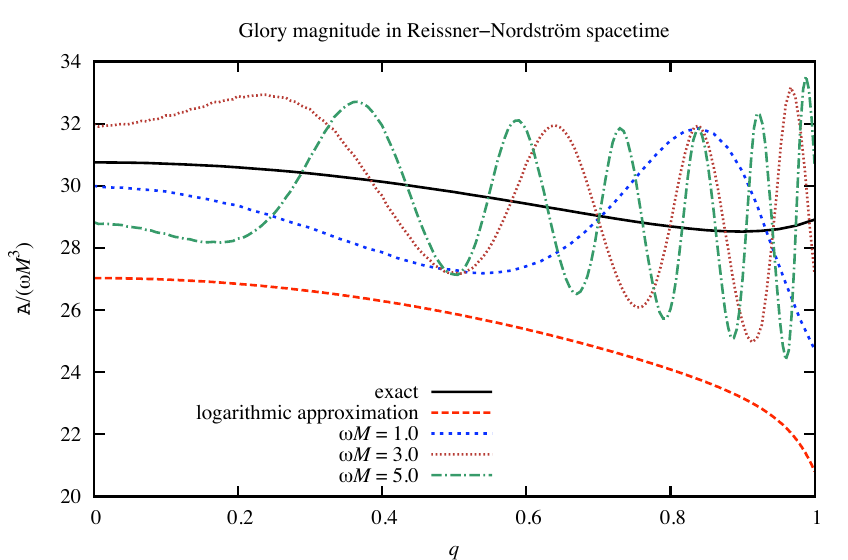}
 \caption{Intensity of the glory peak $\mathcal{A}$ as 
 a function of $q$. 
 The prediction of the logarithmic approximation is compared with 
 the exact solution of the orbital equation.
 We also show the intensity of the glory peak computed numerically via 
 the partial wave method for $\omega M=1.0,3.0, \, {\text {and}} \, 5.0$.}
 \label{fig-glory}
\end{figure}

For near-critical orbits we find that the coefficient $k$ of the elliptic
integrals in Eq.~(\ref{scatt_angle}) behaves as
\begin{equation}
k^2 = 1 - \frac{2 u_c (\bar{u}_3 - \bar{u}_0) \Delta}
{(\bar{u}_3 - u_c)(u_c - \bar{u}_0)} \delta + \mathcal{O}(\delta^2)\,.
\end{equation} 
To derive the logarithmic deflection formula, 
we make use of the approximations for $k\approx 1$
\begin{eqnarray}
K(k) \approx  \frac{1}{2} \ln \left( \frac{16}{1 - k^2} \right)\,, \\
F(z, k) \approx \frac{1}{2} \ln \left( \frac{1+z}{1-z} \right)\,,
\end{eqnarray}
and also $(u_3-u_c)(u_c-u_0) = 2/(Qu_cb_c \Delta)^2$ and $u_3 - u_0 = M \sqrt{1 + y / M} / Q^2$. 
Putting all these elements into Eq.~(\ref{scatt_angle}), we find
\begin{equation}
\Theta(b) \approx - \alpha(q) \ln \left( \frac{b - b_c}{ \beta(q) M} \right)\,,
\label{scatt_angle_approx}
\end{equation} 
where the dimensionless coefficients are
\begin{eqnarray}
\alpha(q) &=& \frac{u_c b_c}{\sqrt{6 - u_c^2 b_c^2}}\,, \label{alpha} \\
\beta(q)   &=& \frac{32 (6 - u_c^2 b_c^2)^3}{u_c^6 b_c^3 M^3 (1 + y/M)} 
\frac{(1-z)^2}{(1+z)^2} e^{-\pi / \alpha(q) } \label{beta} \,.
\end{eqnarray}

The glory formula is
\begin{equation}
\left.\frac{d\sigma_{sc}}{d\Omega}\right|_{\theta\approx\pi} \approx  
\mathcal{A}(q)  \, 
\left[ J_{2s} (b_g \omega \sin \theta) \right]^2 \label{glory-1}\,,
\end{equation}
where the magnitude of the glory peak is given by
\begin{equation}
\mathcal{A}(q) = 2 \pi \omega b_g^2 \left| \frac{d b}{d \theta} \right|_{\theta=\pi} \,.
\label{Aq}
\end{equation}

In Fig.~\ref{bgdbdt} we compare the values of $b_g$ and $|db/d\theta|_{\theta = \pi}$ 
calculated from approximation (\ref{scatt_angle_approx}) with exact values determined 
from numerical integration. It shows clearly that the estimate of $b_g$ found 
from (\ref{scatt_angle_approx}) is significantly more accurate than the corresponding 
estimate of its derivative with respect to $\theta$.

The magnitude of the glory peak obtained using the logarithmic approximation 
[Eqs.~(\ref{scatt_angle_approx})--(\ref{beta}) and~(\ref{Aq})] 
is plotted in Fig. \ref{fig-glory}, and it can be seen that it decreases with $q$.
The logarithmic scattering results suggest that the glory magnitude 
for $q=1$ should be significantly smaller than for $q=0$. 
In Fig.~\ref{fig-glory} we also show the values of $\mathcal{A}(q)$ 
obtained by solving the orbital equation (\ref{Geo1}) numerically. 
It is interesting that these two approaches disagree significantly 
near $q=1$
(the curve obtained using the orbital equation goes up whereas the 
logarithmic approximation curve goes down).
It is clear that the exact solution does not agree with the logarithmic 
approximation for the glory scattering. For instance, for the Schwarzschild case,
the logarithmic approximation gives~\cite{Darwin} 
$\mathcal{A}(q=0) = 27.029 \omega M^3$ 
whereas the exact value is $\mathcal{A}(q=0) =30.752 \omega M^3$. We
find $b_g = 5.346635 M$ and $|db/d\theta|_{\theta=\pi} = 0.150483 M$
for the
logarithmic approximation, compared with 
$b_g = 5.356959 M$ and $|db/d\theta|_{\theta=\pi} = 0.170554M$, 
obtained numerically. 
As we can see from Fig.~\ref{bgdbdt}, most of the error in the 
logarithmic approximation comes from the derivative of $b$ 
with respect to $\theta$.
The values of $\mathcal{A}(q)$ obtained via the partial wave method 
(cf. Sec.~\ref{pwmscs}) for 
$\omega M=1.0,3.0, \, {\text {and}} \, 5.0$ 
are also shown in Fig.~\ref{fig-glory}.
We see that they oscillate around the semi-classical result obtained 
using the orbital equation (\ref{Geo1}).

Finally we note that the glory approximation (\ref{Gen glory}) may be 
improved by including the contribution from geodesics passing more than 
once around the black hole (i.e. through angles $3\pi$, $5\pi$, etc.)~\cite{FHM}. 
Higher-order contributions of this kind will be suppressed by successive 
factors of $e^{-2\pi / \alpha(q)}$. Since the largest value of $\alpha(q)$ 
is $\alpha(1) \approx 1.4142$, subsequent contributions will be suppressed 
by at least $e^{-2\pi / 1.4142} \approx 1.2 \times 10^{-2}$. 
We neglect these contributions here, although elsewhere it was 
shown~\cite{Dolan-2008} that adding the second-order contribution may 
improve the approximation slightly. 

\end{appendix}

\end{document}